\documentclass{article}
\usepackage[letterpaper,top=2cm,bottom=2cm,left=3cm,right=3cm,marginparwidth=1.75cm]{geometry}
\usepackage{amsmath, amsfonts, MnSymbol}
\usepackage{lscape}
\usepackage{graphicx}
\usepackage[numbers,sort,square]{natbib}

\title{Accounting for the absence of anomalous microwave emission in the M 31 halo}

\author{Francesco De Paolis$^{1, 2, 3, \dagger}$, Faryal Naseem$^{4, \ddagger}$, Noraiz Tahir $^{4,*}$ \\
$^1$Department of Mathematics and Physics ``Ennio De Giorgi'', \\ University of Salento, Via per Arnesano, \\ I-73100  Lecce, Italy. \\
$^2$INFN, Sezione di Lecce, Via per Arnesano, I-73100 Lecce, Italy.\\
$^3$INAF, Sezione di Lecce, Via per Arnesano, I-73100 Lecce, Italy.\\
$^4$Department of Physics and Astronomy, \\ School of Natural Sciences, National University of Sciences and Technology, \\ H-12, 44000, Islamabad, Pakistan. \\
$^\dagger$ francesco.depaolis@le.infn.it\\
$^\ddagger$ faryalnaseem789@gmail.com\\
$^*$ noraiz.tahir@sns.nust.edu.pk}

\date{Submitted:08-04-2025; Accepted: 14-10-2025}
\begin{document}
\maketitle

\begin{abstract}
The discovery of a temperature asymmetry in the cosmic microwave background (CMB) data towards various galaxies has opened a window for a deeper comprehension of galactic halos. A crucial step forward is that of estimating the fraction of missing baryons in the halos, but it relies on understanding the real cause of the observed CMB temperature asymmetry since many effects might give a non-negligible contribution. Here, we analyzed the contribution played by the anomalous microwave emission (AME) from halo dust grains in the halo of the M 31 galaxy. Assuming either amorphous carbon and silicates dust grains with size ranging from $0.01~\mu$m to about $0.3~\mu$m and mass in the range $10^{-14} - 10^{-13}$ g, we estimated the total mass, distribution, and diffuse emission in the $100\,\mu$m band of the Infrared Astronomical Satellite (IRAS). Then, we estimated the temperature asymmetry induced by the rotation of the M 31 halo and compared the obtained values with the \textit{Planck}'s SMICA-processed data. We find that the AME cannot account for the measured CMB temperature asymmetry, with its contribution constrained to $\lesssim 7\%$, thereby indicating that additional physical mechanisms must be responsible for the observed signal.
\end{abstract}


\section{Introduction \label{intro}}
The analysis of CMB data from the WMAP and \textit{Planck} has revealed a notable temperature asymmetry—ranging between about $50$ and $80\,\mu$K towards the halos of several nearby edge-on spiral galaxies, including M 31, M 33, M 81, M 82, Centaurus A, M 104, and M 90 \citep{de2014planck, de2015planck, gurzadyan2015planck, de2016planck, gurzadyan2018messier, de2019rotating, depaolis2024m90}. This asymmetry extends up to galactocentric distances  $\sim 100$ kpc and is observed to be frequency-independent, strongly indicating a Doppler-induced origin due to the galactic halos rotation. Several physical mechanisms have been proposed to explain this phenomenon: $(i)$ the rotational kinetic Sunyaev-Zeldovich (rkSZ) effect arising from the motion of hot ionized gas in rotating halos \citep{zorillamatilla2020}; $(ii)$ synchrotron radiation produced by fast-moving cosmic ray electrons in magnetic fields \citep{2000AA...362..151D}; $(iii)$ free-free (bremsstrahlung) emission from hot electrons scattering off ions \citep{sun2010galactic}; $(iv)$ anomalous microwave emission (AME) from spinning dust grains present in the interstellar medium, including in galactic halos \citep{Erickson1957, Hoyle1970, kogut1996microwave, Leitch2013TheDO}; and $(v)$ thermal emission from cold or virialized clouds in the galactic halo, which may contain both gas and dust components \citep{1995AA...299..647D, qadir2019virial, tahir2019constraining, planck2013thermal}.
 
Currently, option ($v$) seems to provide the most likely explanation of the observed temperature asymmetry, but to really trust this option, a precise estimate of the contribution from the other effects has to be provided. The rkSZ effect may be relevant on the scale of galaxy clusters where the hot diffuse gas can reach temperatures exceeding a few times $10^7$ K and sizes can be of a few Mpc \citep{cooray2002,altamura2023}, on the scale of galactic halos it certainly plays a less important role \citep{zorillamatilla2020}. In the case of M 31 the hot diffuse gas is primarily confined to the inner regions, with X-ray observations indicating its dominance within a radius of approximately $2.5$ kpc, where it extends vertically above the disk and traces the bulge emission \citep{li2007}. In contrast, the stellar halo of M31 becomes increasingly prominent at larger galactocentric radii and has been detected out to projected distances exceeding $175$ kpc, corresponding to about two-thirds of the galaxy's virial radius \citep{gilbert2014}. 

Recently, \cite{tahir2022} estimated the rkSZ induced temperature asymmetry, the value of the estimated $\Delta T$ came out to be in range $1.16 \times 10^{-1}~\mu$K - $7.89 \times 10^{-2}~\mu$K within a galactocentric distance $21.4 - 108.3$ kpc. However the observed temperature asymmetry in {\it Planck}'s SMICA (Spectral Matching Independent Component Analysis) data towards M 31 halo is in the range $2.66~\mu$K - $21.4~\mu$K. That means that the rkSZ effect certainly plays a minor role since it can contribute only up to about $4.3\%$ to the observed temperature asymmetry towards the M 31 halo.

Considering effect ($iv$), AME from dust grains remains one of the less well understood galactic foregrounds. Its phenomenological modeling has been advanced by \cite{ali2013, zang2025}, while proposed carriers (ultra-small grains, PAHs, magnetic nanoparticles) continue to be explored \citep{hensley2016}. AME has now been robustly detected in Galactic and extragalactic environments by  \cite{PlanckAME2015, Fernandez2023, murphy2010}. 

Specifically for M31, AME has been previously investigated in the central and disk regions by \citet{PlanckAME2015}, who reported tentative excess microwave emission possibly associated with spinning dust.  \citet{Tibbs2012} and \citet{Browne2019} analyzed localized AME features in M31. Previous studies of AME have typically focused on resolved regions within galaxies, such as star-forming areas or the central bulge. In contrast, integrated (whole-galaxy) analyses are less common. An example of such a study is the detection of AME in M31 using QUIJOTE-MFI data \citep{Fernandez2024}. Additionally, upper limits for AME in several entire galaxies have been reported in \citep{Bianchi2022}. These whole-galaxy investigations provide valuable constraints on AME on global scales, complementing the more localized studies. Observational correlations between AME and thermal dust emission have been found in the Milky Way and external galaxies, including M 31 \citep{PlanckAME2015, Tibbs2012,dickinson2018}, supporting a common origin. Both are associated to interstellar dust grains and both can allow tracing dust-rich regions. While thermal dust emits predominantly at higher frequencies ($\gtrsim 100$ GHz) and is generally associated to large grains in equilibrium with the interstellar radiation field, AME is believed to arise from the electric dipole radiation of rapidly rotating very small grains, particularly polycyclic aromatic hydrocarbons (PAHs), emitting in the $10$ – $60$ GHz range (see, e.g., \citealt{Ysard2022}). Observational correlations between AME and thermal dust emission have been found \citep{cepeda2020, Fernandez2023}.

In contrast to earlier studies focusing on the M 31 disk and bulge, the present paper aims investigating the contribution of AME in the extended halo of the M 31 galaxy, where the observed temperature asymmetry reaches several tens of $\mu$K across all {\it Planck} bands, extending up to about $10^\circ$ from the galaxy center, corresponding to about 135 kpc. The dust mass $M_d$ in the M 31 galaxy derived from observations amounts to about $10^7~M_{\odot}$ - $10^8~M_{\odot}$  \citep{2003ARA&A..41..241D, 2018ApJ...854...36I} and the more robust direct estimate within $25$ kpc is $M_d(R\leq 25 \,kpc)\simeq 5.4\times 10^7~M_{\odot}$, while the estimated gas mass within the same galactocentric distance amounts to about $M_g\simeq 6.7\times 10^9~M_{\odot}$ \citep{tamm2012}. Then, the dust-to-gas ratio could vary across the galaxy decreasing from the center outward, should be approximately $0.008$ while the ratio of the dust mass with respect to the full M 31 virial mass (about $10^{12}~M_{\odot}$ up to $200$ kpc) would be $\simeq 5.4\times 10^{-5}$ or slightly higher.  Note that an estimate of the baryons amount in the M 31 circumgalactic medium within its virial radius was  attempted recently using $\gamma$-ray observations \citep{zhang2021measuring} obtaining about a few times $10^{10}\,M_{\odot}$. One could expect that AME might not be important in galactic halos due to the limited amount of dust that should be present in those  regions. Therefore, most likely, AME should not allow accounting for the temperature asymmetry in the microwaves detected by \textit{Planck} satellite \citep{de2014planck}. However, it is worth investigating this issue since the detection of AME emission in galactic halos should allow getting important information about the amount and distribution of spinning small dust grains in those regions.

Dust grains in galactic halos are thought to originate in the disk and transported outward by radiation pressure. Their diffusion into the halo depends on factors such as the galaxy’s mass, the temperature of the hot-gas halo, and the radiation pressure coefficients. Lighter grains like graphite are more efficiently accelerated than heavier silicate grains \citep{60}. However, not all grains reach the halo due to the counteracting effects of strong radiation pressure and galactic magnetic fields \citep{hirashita2020dust}. In this study, we assume that M 31 halo dust consists primarily of amorphous carbon and silicates. In Section \ref{amemodel}, we model the spatial distribution of halo dust grains, estimate their total infrared (IR) emission and temperature using IRAS data within 135 kpc, and use this to constrain the grain size. In Section \ref{deltaT}, we calculate the temperature asymmetry these grains could induce and compare it with the SMICA-processed \textit{Planck} data \citep{planck2018SMICA}. Our main findings are discussed in Section \ref{results}.

\section{Distribution of halo dust grains \label{amemodel}}
It has been proposed that AME is caused by very small dust grains rapidly spinning, emitting radiation at microwave frequencies \citep{1994ApJ...427..155F, draine1998electric}.  As a result, AME can be distinguished from other emission processes, such as synchrotron radiation or free-free emission. Indeed, the AME spectral index rises at low frequency and then falls of and is not well described by a single power-law, synchrotron radiation is characterized by a single power-law index $\simeq 0.7-1.0$ while free-free emission has a flat or slightly declining spectrum with spectral index $\simeq - 0.1$. Moreover, polarization in AME is negligible with respect to synchrotron emission \citep{Gonzalez2025}. AME has been detected in the \textit{Planck} $10$ – $60$ GHz frequency band, which exhibits a strong link to far infrared (FIR) radiation \citep{PlanckAME2015,Fernandez2023} that is typically associated with the thermal emission from dust grains \citep{kogut1996microwave, leitch1997anomalous}. 

\begin{figure}
    \centering
    \includegraphics[width=0.8\linewidth]{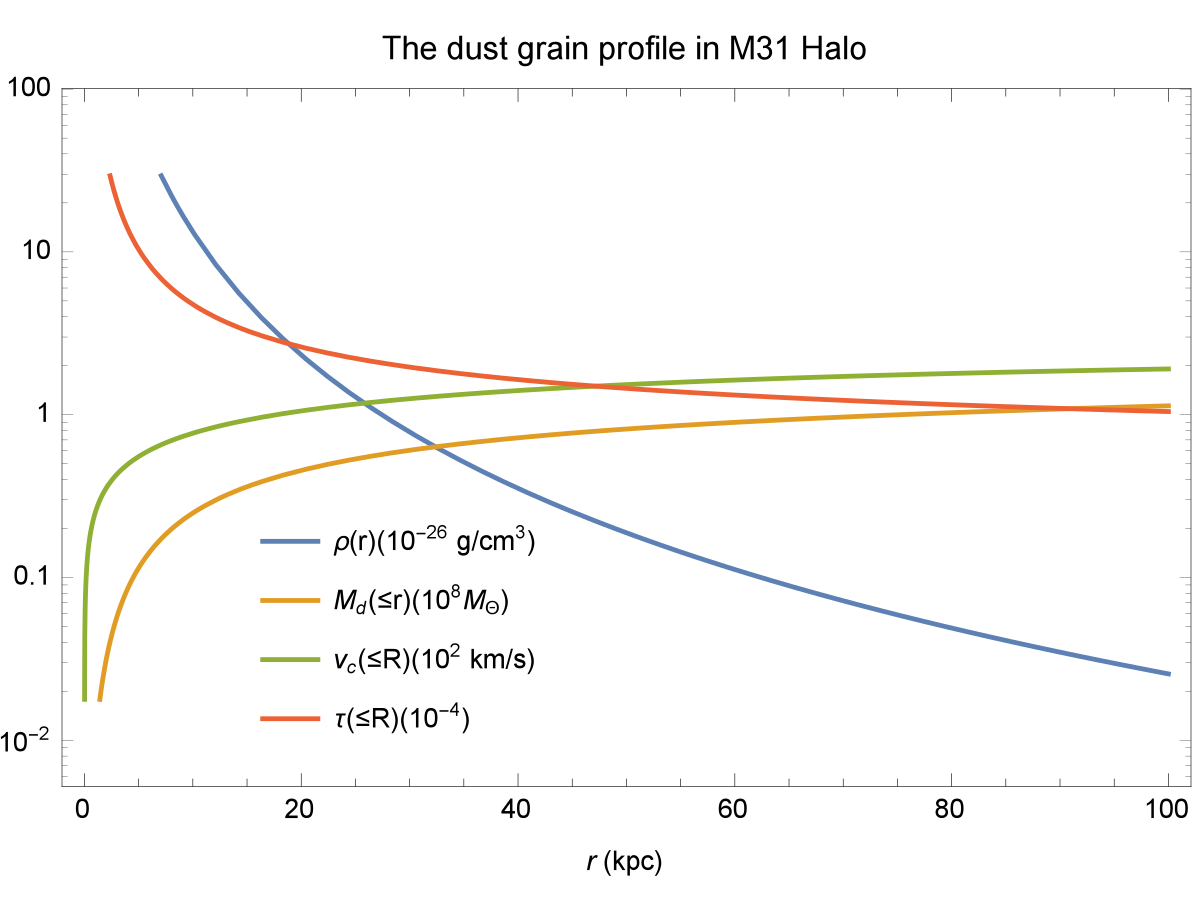}
   \vspace{-0.8cm}
    \caption{M 31 halo profiles of the dust grain density  (blue curve), dust mass (orange curve), circular velocity (green curve), and optical depth (red curve).}
    \label{mergedall}
\end{figure}

To estimate the AME contribution in the CMB data towards the M 31 halo we follow \cite{amekhyan2019role} and  assume that halo dust is distributed according to the Navarro-Frenk-White (NFW) density profile \citep{navarro1997universal}, i.e. 
\begin{equation}
    \rho(r) = \displaystyle{\frac{f~\rho_c}{\left(r/r_c\right) \left(1 + r/r_c\right)^2}}\, ,
    \label{NFW}
\end{equation}
where $f$ is the dust fraction with respect to the halo dark matter, $r$ is the galactocentric radial distance, $\rho_c = 1.74 \times 10^{-2}~M_{\odot}\,{\rm pc^{-3}}$ is the dark matter central density, and $r_c = 12.4$ kpc is the M 31 halo core radius. Note that in our toy model we assume that the fraction $f$ is constant throughout the whole M 31 halo, leaving to an eventual deeper study the generalization of $f$ to $f(r)$. The dust mass within the projected radial distance $R$ (with respect to the M 31 rotation axis) is estimated by using the relation
\begin{equation}
    M_d(\leq R)= \int_0^R 4 \pi r^2 \rho(r) dr\,. 
    \label{mass}
\end{equation}
 The dust mass density and the dust mass profiles in the M 31 halo are shown in Fig. \ref{mergedall}. Setting, $f=3.14 \times 10^{-4}$ we obtain a consistent value for the halo dust mass obtained by \citealt{draine2013andromeda} within about $25$ kpc. Following \cite{tahir2022} the circular velocity of the halo dust grains can be estimated by  
\begin{equation}
    v_c(\leq R) = \sqrt{\alpha\left[\ln \left(1+R/R_{vir}\right)-\left(R/R_{vir}\right)\left(1+R/R_{vir}\right)^{-1}\right]}\, ,
    \label{generalvc}
\end{equation}
where $\alpha= (4 \pi G r_c^2 \rho_c R_{vir})/R$, $G$ is  Newton's gravitational constant, and $R_{vir} \simeq 200$ kpc is the M 31 virial radius. The radial profile of the dust rotational velocity in the M 31 halo obtained by eq. (\ref{generalvc}) is shown in Fig. \ref{mergedall} and the asymptotic value of the circular velocity is found to be $\simeq 230$ km\,s$^{-1}$. Note, however, that the values of the circular velocity given by eq. (\ref{generalvc}) have to be considered as upper limits to the actual bulk rotational speed of the M 31 halo, since that equation derives from the assumption that the centrifugal equilibrium holds, an assumption that might not apply, in general, to galactic halos.  

\section{Infrared emission from halo dust towards M 31 \label{dustemission}}

Observations of the M 31 galaxy have revealed the presence of cold dust extending beyond the central regions into the outer disk and halo. Infrared observations, particularly by IRAS and ISO (Infrared Space Observatory), have allowed the detection of dust with a temperature of about $16$ K in the M 31 outer regions \citep{haas1998}. These authors mapped the cold dust in M 31 using ISO $175\,\mu$m observations, revealing a dominant dust ring at approximately $10$ kpc and a fainter outer ring at about $14$ kpc. They found that the bulk dust temperature is about $16$ K, significantly colder than the $21$--$22$ K estimated from IRAS data within $25$ kpc. Spectral energy distribution analysis confirmed the presence of two dust components: a cold dust population (at  a temperature of about $16$ K) and a warmer component (with $T \simeq 45$ K). 

The aim of this section is to use IRAS dust excitation data within $10$ deg ($\simeq 135$ kpc) from the M 31 center to estimate the radiative flux from the halo dust grains and  map the dust temperature in the M 31 halo. We made use of the dust extinction service of IRAS, which is an interactive service that uses the maps of Schlegel, Finkbeiner, and Davis (SFD) to return extinction along the line of sight through the considered galaxy for $100 \,\mu$m. In this analysis, the dust temperature was not directly computed from the IRAS $100\,\mu$m band alone; rather, it was inferred through the use of the temperature-corrected dust emission map provided by the  SFD98 model \citep{Schlegel1998}. We combined the SFD98 model high-resolution at $100\,\mu$m IRAS maps with temperature estimates derived from the COBE/DIRBE $100\,\mu$m and $240\,\mu$m bands, under the assumption that dust emission follows a modified blackbody spectrum with a fixed emissivity index $\beta = 2$. This allowed the construction of a full-sky temperature-corrected dust map that traces dust column density more accurately. The corrections also involved careful subtraction of background components, including zodiacal light which is modeled and removed using DIRBE's seasonal variation, the cosmic infrared background (CIB) which is estimated from high-latitude sky regions, and instrumental systematics. Since the IRAS data alone lacked absolute calibration, we used the SFD98 model to recalibrate it using DIRBE data in order to ensure consistency.

\begin{figure*}
    \centering
    \includegraphics[width=1.0 \linewidth]{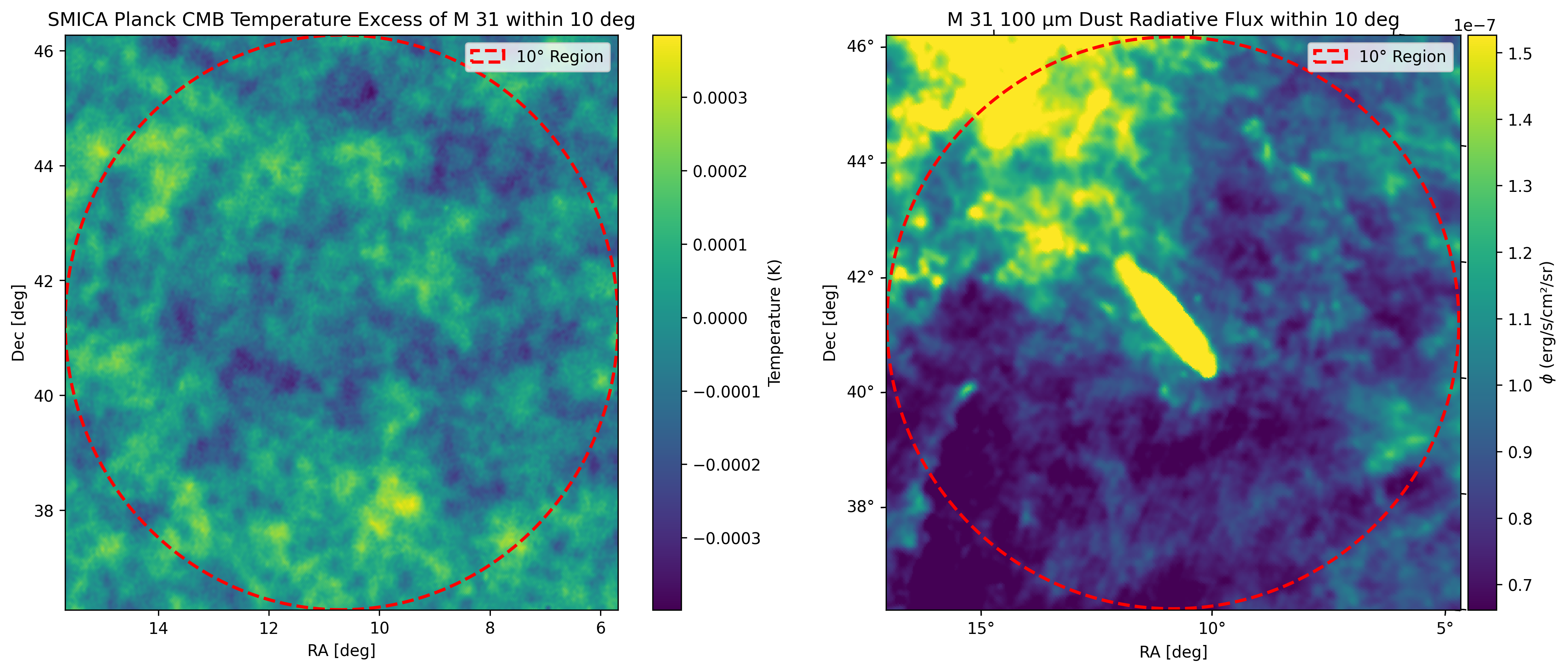}
    \vspace{-0.4cm}
    \caption{Left Panel: \textit{Planck}'s SMICA map of the temperature excess within 135 kpc towards M 31. Right Panel: IRAS dust excitation map at  $100\,\mu$m within 135 kpc.}
    \label{tempearturemaps}
\end{figure*}

The average temperature of the M 31 halo dust grain from the analysis came out to be $17.7 \pm 0.2$ K. The radiative flux was estimated from the temperature-corrected IRAS $100\,\mu$m dust emission map provided by the SFD98 model and is shown in the right panel of Fig. \ref{tempearturemaps} while in the left panel of the same figure the {\it Planck} SMICA map of the temperature excess within the 135 kpc region is given. The radiative flux from the M 31 halo came out to be in the range  $7.08 \times 10^{-8} - 1.54 \times 10^{-7} \, {\rm erg\, s^{-1}\,cm^{-2}\,sr^{-1}}$ within about $10$ deg from the M 31 center.

We model the dust grain emission using the observations towards the M 31 halo by \cite{1984ApJ...278L..59H, hirashita2020dust, 2020MNRAS.492.5052Y} in mid and FIR wavelengths, and assuming that the M 31 halo contains carbon and silicates grains with sizes $a$  from $0.01\,\mu$m to $0.3 \,\mu$m and mass  $\simeq 10^{-14} - 10^{-13}$ g. The dust grain optical depth  $\tau$  within the M 31 galactocentric distance $R$ can be estimated as  (see \citealt{2019IJMPD..2840016A})
\begin{equation}
    \tau(\leq R) = \frac{M_d (\leq R)\kappa}{D^2 \Omega}. 
    \label{dustmass}
\end{equation}
Here, $D = 744$ kpc is the M 31 distance, $\Omega$ is the solid angle corresponding to the radial distance $R$, and $\kappa$ is the absorption cross section of the dust grains (in units of ${\rm cm^2 \,g^{-1}}$), which is given by \cite{amekhyan2019role}
\begin{equation}
    \kappa \simeq \int_{\nu_i}^{\nu_f} 0.1 \left(\frac{\nu}{10^3\, {\rm GHz}}\right)^{\beta} d\nu ~.
\end{equation}
 Here, $\beta \simeq 1$ is the power-law index for amorphous molecules \citep{2003ARA&A..41..241D}, and $\nu_{i}=300$ GHz, and $\nu_f = 20$ THz are the lower and upper FIR frequencies of IRAS \citep{1984ApJ...278L..59H}. The obtained optical depth, up to 135 kpc from the M 31 center, is shown in Fig. \ref{mergedall}. As one can see, the obtained values in the FIR band are in agreement with those reported in \cite{amekhyan2019role}. 
The  density $\Phi ( a, T_d, \leq R)$, in units of {\rm erg\, s$^{-1}$ cm$^{-2}$\, sr$^{-1}$}, can be estimated by using the relation
\begin{align}
    \Phi( a, T_d, \leq R) =\frac{4\pi ~ B(T_d)}{m_{gr}} \int_0^R \tau(\leq R) \rho(R) R^2 dR \, . 
    \label{density}
\end{align}
Here, $m_{gr} = (4/3) \pi a^3 \rho_{gr}$ is the mass of a single dust grain with density $\rho_{gr}$, that is about 1 g\, cm$^{-3}$ for the case of either amorphous carbon or silicates, and radius $a$. The blackbody spectral radiance $B(\nu, T_d)$ integrated over the frequency is given by
\begin{equation}
    B(T_d) = \int_{\nu_i}^{\nu_f} \frac{2 h \nu^3}{c^2} \frac{1}{\exp\left( \frac{h \nu}{k_B T_d}\right)- 1} d\nu,
    \label{blackbody}
\end{equation}
where $h$ is Planck's constant, $k_B$ is the Boltzmann constant, $c$ is the speed of light, and $T_d$ is the dust grain temperature. The obtained values of the radiative flux $\Phi$ are listed in Table \ref{tabphi} (column 2) for a range of grain sizes (column 1). These results align closely with estimates derived from IRAS dust excitation data and \cite{2019ApJ...877L..31B} analysis. It should be noted that  \cite{2019ApJ...877L..31B} do not explicitly report radiative flux values and their analysis is limited to the M 31 disk extension. Indeed, their study was confined to an elliptical region of $91.5' \times 59.5'$, which turns out to be approximately $42 \times 27$ kpc at the distance of M 31. This region is comparable only with the innermost radial region that we consider, i.e. that with radius of $21.4$ kpc.  The average radiative flux within $21.4$ kpc region by \cite{2019ApJ...877L..31B} in all the \textit{Planck} bands came out to be $7.91 \times 10^{-7} \, {\rm  erg\, s^{-1}\, cm^{-2}\, sr^{-1}}$. The average flux value we obtained for the various considered grain sizes is of the same order of magnitude as the flux reported by \citet{2019ApJ...877L..31B}. However, we note that the modeled fluxes span approximately one order of magnitude depending on the grain size. In particular, smaller grains in the range of $0.01 - 0.06\,\mu$m yield fluxes that best reproduce the observed value, consistent with expectations for AME being dominated by very small spinning dust grains (see Table \ref{tabphi}). As a result, this consistency in the underlying dust emission trends across both studies reinforces the interpretation that AME from spinning dust grains may be a viable mechanism across multiple galactic scales.

\begin{table}[ht]
\caption{The estimated radiative flux within 135 kpc from M 31 halo dust grains.}      
\label{tab1}    
\centering                            
\begin{tabular}{c c}      
\hline\hline     
\vspace{0.2 cm}
Dust Grain Size & Radiative Flux\\
$a$ ($\mu$m) & $\Phi$ ($10^{-7}~{\rm erg\, s^{-1}\, cm^{-2}\, sr^{-1}}$)   \\ 
\hline
0.01   & $9.90$  \\
0.06   & $5.37$  \\
0.11   & $4.41$\\
0.16   & $3.90$  \\
0.21   & $2.92$ \\
0.26   & $0.81$ \\
0.36   & $0.98$ \\
\hline 
\end{tabular}
\vspace{0.2cm}

\textbf{Note:} The estimated radiative flux within 135 kpc from M 31 halo dust grains at $T_d\simeq 17$ K for various dust grain sizes $a$ (first column) is given in the second column.

\label{tabphi}
\end{table}
\section{Temperature asymmetry contribution by AME \label{deltaT}}
In the previous Section we constrained the M 31 halo dust parameters. The temperature asymmetry in the CMB caused by AME from the rotation of the M 31 halo with cicrular speed $v_c$ can be estimated by using the relation 
\begin{equation}
    \Delta T \left(R, \phi, a\right)=\frac{T_d \sigma(a)}{c}\int_{los} n_d(r, a)v_c(R) \cos\phi \sin i~dl,
    \label{deltat}
\end{equation}
where, $n_d(r, a) = \rho(r)/m_{gr}(a)$ is the number density of the halo dust grains, $\phi$ is the azimuthal angle, $i\simeq 77.5^\circ$ is the inclination angle of the M 31 galaxy, $\sigma$ is the grain cross section, and the integral is performed along the line of sight (los).
In Table \ref{tabdeltaT} we give the values of the estimated temperature asymmetry induced by the AME effect (in the third column) for various grain sizes given in the first column, and compare them with the observed value of the temperature asymmetry given in the fourth column. 
The observed temperature asymmetry  corresponds to the foreground corrected SMICA-processed data. We note that the SMICA data were chosen because they are less contaminated from unresolved sources and other foregrounds at small angular scales with respect to the other available \textit{Planck} maps 
\citep{planck2014,planck2018SMICA}. As one can see, the estimated values of $\Delta T$ due to AME are much lower with respect to the asymmetry detected on CMB data ranging between 50-80\, $\mu$K. Hence, we can safely conclude that AME cannot account for the observed temperature asymmetry towards the M 31 halo.

Moreover, in order to establish the relative importance of the AME induced temperature asymmetry in the inner region of the  M 31 galaxy with respect to the halo, we considered  in particular the $ 1- 10$ kpc region. The estimated AME induced temperature excess $\Delta T_{est}$ turned out to be in the range $0.62-2.25 \, \mu$K, for $a = 0.01\,\mu$m down to $\Delta T_{est} = 0.25-1.24\,\mu$K for $a = 0.36\,\mu$m. It is clearly seen, by comparing these values with those reported in the third column of Table \ref{tabdeltaT}  that the AME induced temperature asymmetry is more significant in the inner regions of the M 31 galaxy and turns out to play a minor role in the regions beyond about $20$ kpc of the M 31 galaxy.

\begin{table}[]
\caption{Estimated temperature asymmetry induced by AME from dust grains in M 31 halo.}       
\centering                                    
\begin{tabular}{c c c c}        
\hline\hline                      
Radial & Observed & Halo Dust  & Estimated \\
Region & $\Delta T_{obs}$ & Grain Size  & $\Delta T_{est}$  \\
kpc & $\mu$K & $a$ ($\mu$m)    & ($10^{-2}\, \mu$K) \\ 
\hline
21.4 & 2.66 & 0.01  & 19.79    \\
    & & 0.06 & 13.01 \\
    &  & 0.11 & 11.69  \\
    &  & 0.16 & 11.62 \\
    & & 0.21 & 8.86 \\
    &  & 0.26 & 7.15 \\
    &  & 0.36 & 5.99 \\
31.1 & 6.44 & 0.01  & 19.84    \\
    & & 0.06 & 13.25 \\
    &  & 0.11 & 11.94  \\
    &  & 0.16 & 11.52 \\
    & & 0.21 & 8.95 \\
    &  & 0.26 & 7.22 \\
    &  & 0.36 & 6.02 \\
41.4 & 17.60 & 0.01  &19.91    \\
    & & 0.06 & 13.42 \\
    &  & 0.11 & 11.71  \\
    &  & 0.16 & 11.72 \\
    & & 0.21 & 9.02 \\
    &  & 0.26 & 7.25 \\
    &  & 0.36 & 6.14 \\
51.8 & 39.20 & 0.01  &20.35    \\
    & & 0.06 & 13.56 \\
    &  & 0.11 & 11.02  \\
    &  & 0.16 & 11.00 \\
    & & 0.21 & 9.12 \\
    &  & 0.26 & 8.08 \\
    &  & 0.36 & 6.23 \\    
77.8 & 21.70 & 0.01  &20.21    \\
    & & 0.06 & 13.68 \\
    &  & 0.11 & 10.81  \\
    &  & 0.16 & 9.88 \\
    & & 0.21 & 9.12 \\
    &  & 0.26 & 8.10 \\
    &  & 0.36 & 6.48 \\ 
103.8 & 39.20 & 0.01  &20.32    \\
    & & 0.06 & 13.78 \\
    &  & 0.11 & 10.35 \\
    &  & 0.16 & 9.91 \\
    & & 0.21 & 8.72 \\
    &  & 0.26 & 7.12 \\
    &  & 0.36 & 6.50 \\     

                       \hline
\end{tabular}
\vspace{0.2cm}

\textbf{Note:} Estimated temperature asymmetry induced by AME from dust grains (column 4) for various grain sizes $a$ (column 3) within the radial region (column 1). A dust temperature $T_d \simeq 17$ K was assumed. The SMICA observed temperature asymmetry is also given in column 2.
\label{tabdeltaT}
\end{table}

Before closing this section, we mention that we analyzed the dependence of the radiative flux and $\Delta T$ values on the dust grain temperature. We considered  dust temperatures from $30$ K down to about $2.5$ K. It was seen that decreasing the temperature, the fraction $f$ increases and, as a result, $\Delta T$ also increases. For example, for $30$ K dust grains, we have $f \simeq 10^{-5}$ and $\Phi$ in the range between $4.65 \times 10^{-7}~~ {\rm erg\,s^{-1}\, cm^{-2}\, sr^{-1}}$ and $6.45 \times 10^{-8}~~ {\rm erg\,s^{-1}\, cm^{-2}\, sr^{-1}}$, for grain sizes between $0.01 \mu$m and $0.3 \mu$m. Consequently, the temperature asymmetry $\Delta T$ came out to be about $4.65 \times  10^{-3}\, \mu$K at $21.4$ kpc to $5.65 \times 10^{-2}\, \mu$K at $103.8$ kpc. Similarly, for cold dust grain sizes between $0.01 \mu$m and $0.3 \mu$m at $T_d = 2.5$ K we get $f\approx 10^{-1}$ while the radiative  flux $\Phi$ came out to be in the range $2.31\times 10^{-7}-4.78 \times 10^{-7}~~ {\rm erg\,s^{-1}\, cm^{-2}\, sr^{-1}}$. Consequently, $\Delta T$ came out to be about $1.89 \times 10^{-2} \, \mu$K at $21.4$ kpc to $20.45 \times 10^{-2} \, \mu$K at $103.8$ kpc. Therefore, the dust temperature seems to play a rather relevant role in the AME induced temperature asymmetry. Nevertheless, even under these extreme cold-dust conditions, the AME contribution remains far below the observed level of CMB temperature asymmetry, being limited to $\lesssim 7\%$ of the measured effect.
\section{Results and conclusions \label{results}}
The observed temperature asymmetry in CMB data towards various edge-on spirals has provided a novel approach to investigate and  map halo dynamics, and study either the disk and halo rotation, even if the true cause of this asymmetry is still unknown. Several theoretical effects could contribute, including the rkSZ effect, which has been analyzed for the halos of M 31 and M 90 \citep{tahir2022, depaolis2024m90}. However, it has been found that this effect is minor, accounting for only about $1\%$ of the observed temperature asymmetry. Here, we explored the potential contribution of the AME from halo dust grains to the CMB temperature anisotropy towards the M 31 halo, assuming that the dust distribution follows the NFW profile and the dust circular velocity profile is given by eq.  (\ref{generalvc}).

In order to estimate the temperature asymmetry $\Delta T$ caused by dust grains, we constrained the grain size and temperature of the M 31 halo dust. We used the IRAS dust excitation data in the $100\,\mu$m band within $10$ deg to estimate the temperature profile of the M 31 halo dust and the radiative flux $\Phi$. These quantities came out to be $T_d\approx 17$ K and $\Phi \approx 10^{-8}~{\rm erg\,s^{-1}\, cm^{-2}\, sr^{-1}}$. The optical depth $\tau$ of the dust grains in the FIR band turned out to be in the range $10^{-3} - 10^{-4}$ within about 135 kpc, consistent with that obtained by \cite{amekhyan2019role} in the microwave band. 

We then estimated the expected temperature asymmetry $\Delta T_{est}$ caused by AME as a function of the projected radial distance of the M 31 halo by using eq. (\ref{deltat}). We found that the AME contribution from halo dust at $17$ K is variable from $0.3\%$ to about 7$\%$ at most, and can be considered negligible with respect to the measured $\Delta T_{obs}$ (see Table \ref{tabdeltaT}). 

Moreover, when comparing the central ($<20$ kpc) and outer ($>20$ kpc) regions, the AME signal is found to drop significantly. This behaviour closely follows the underlying dust density profile, which is expected since the AME distribution was modelled on it. In fact, this steep decline indicates that spinning dust emission is unlikely to be a significant component of the microwave sky beyond a certain galactocentric distance. A similar trend was reported by \cite{Fernandez2024}, who measured a reduction in the AME fraction when comparing two aperture sizes centered on M31, one of $91.5' \times 59.5'$ and a larger one of $100' \times 70'$. Although the resolution of their data was not sufficient to consider the two regions fully independent, the larger aperture showed a weaker AME contribution relative to other emission mechanisms. Taken together, these findings support the conclusion that AME is strongly concentrated in the inner region of M31. More importantly, our quantitative analysis shows that, even under the most conservative assumptions, AME contributes at most $\lesssim 7\%$ of the observed CMB temperature asymmetry.

On the other hand, it was shown previously \citep{tahir2022} that the rkSZ from M 31 halo hot gas contribute less than about $1\%$ to the observed temperature asymmetry in {\it Planck} data. Both these findings make cold/virial clouds a stronger candidate than before to account  for the temperature asymmetry revealed towards the M 31 halo.

\section*{Acknowledgments}
FDP acknowledges partial support of the INFN Projects
TAsP (Theoretical Astroparticle Physics) and EUCLID. NT would like to acknowledge the support by the National University of Sciences and Technology (NUST) Project no. NUST-24-41-78. Prof. Asghar Qadir and the anonymous referee are also deeply acknowledged.

\end{document}